# Chiral phonons of honeycomb-type bilayer Wigner crystals

Dingrui Yang[1], Lingyi Li[2], Na Zhang[3], Hongyi Yu[3,4*]

[1] Keble College, University of Oxford, Oxford, United Kingdom

[2] Samueli School of Engineering, University of California, Irvine, CA 92697, USA

[3] Guangdong Provincial Key Laboratory of Quantum Metrology and Sensing & School of Physics and Astronomy, Sun Yat-Sen University (Zhuhai Campus), Zhuhai 519082, China

[4] State Key Laboratory of Optoelectronic Materials and Technologies, Sun Yat-Sen University (Guangzhou Campus), Guangzhou 510275, China

* E-mail: yuhy33@mail.sysu.edu.cn

**Abstract:** We theoretically investigated the chiral phonons of honeycomb-type bilayer Wigner crystals recently discovered in van der Waals structures of layered transition metal dichalcogenides. These chiral phonons can emerge under the inversion symmetry breaking introduced by an effective mass imbalance between the two layers or a moiré potential in one layer, as well as under the time-reversal symmetry breaking realized by applying a magnetic field. Considering the wide tunability of layered materials, the frequencies and chirality values of phonons can both be tuned by varying the system parameters. These findings suggest that bilayer honeycomb-type Wigner crystals can serve as an exciting new platform for studying chiral phonons.

## I. Introduction

In 1934, Eugene Wigner first predicted the existence of a crystalline phase of electrons, namely the Wigner crystal[1]. He pointed out that when the Coulomb potential between electrons is much greater than their kinetic energy which can be realized under a sufficiently low density, electrons will spontaneously break the continuous translational symmetry and arrange in a crystal-like configuration to minimize their overall energy. Following the prediction, many theoretical and experimental studies about Wigner crystals have been carried out by physicists over the next several decades. In 1979, the presence of Wigner crystals is successfully observed on the surface of liquid helium[2], where the nearest-neighbor separation or wavelength is in the order of μm. Later in GaAs-based semiconductor quantum well systems, strong magnetic fields are used to quench the electron kinetic energy, which thus facilitate the realization of Wigner crystals with much smaller wavelengths (~ 30 nm)[3]. In the past decade, layered semiconducting transition metal dichalcogenides (TMDs) has been proved to be an exciting new platform for studying correlated phenomena[4,5], where the reduced screening of the layered geometry largely enhances the Coulomb interaction between electrons. As a result, signatures for the formation of Wigner crystals without the assistance of magnetic fields have been observed in monolayer TMDs recently, with wavelengths ~ 30 nm[6]. In the true two-dimensional limit, the Wigner crystal forms in a triangular lattice configuration. However, theories have predicted that in systems with two vertically separate electron layers, the formed bilayer Wigner crystals can be in other lattice configurations including the staggered rectangular, staggered rhombic and honeycomb lattice types, etc. [7-9]. Recently in bilayer TMDs without magnetic fields, experiment has observed a series of bilayer Wigner crystals under various doping density ratios between the two layers, which are believed to consist of two interlocked triangular-type Wigner crystals located in opposite layers[10]. Under the ratio 1:1, the bilayer Wigner crystal forms a honeycomb-type lattice, with a wavelength as small as 5 nm due to the enhanced stability by the strong interlayer Coulomb interaction[11]. Besides, the formation of moiré patterns in layered TMDs van der Waals structures can introduce moiré superlattice potentials on the electrons, which can significantly change their correlated properties. A weak moiré potential can renormalize the electron effective mass thus affect the formation of the Wigner crystal. A strong moiré potential can localize the electrons to its minima, which can be then described by an extended Hubbard model. Combined with the strong on-site and inter-site Coulomb interactions, a series of correlated electron crystals including the Mott insulators under integer fillings and generalized Wigner crystals under certain fractional fillings have been detected in experiments[12-22].

Due to the long-range nature of the Coulomb interaction, the low-energy excitations of the above-mentioned electron crystals are given by the collective vibration of the lattice sites (i.e., phonon modes). In previous studies about Wigner crystals, their phonon dispersions have been used to analyze the stability, dynamical and melting properties of the crystal[7,8,23-26]. For various electron crystals in monolayer and moiré-patterned bilayer TMDs (including the triangular-, stripe-, and honeycomb-type lattice configurations), our previous work has also studied their phonon properties[27]. In these different electron crystals, the honeycomb-type is particularly interesting as it can host chiral phonons with finite angular momenta and large Berry curvatures[28]. These chiral phonons can give rise to the phonon Hall effect and have potential applications in novel phononic devices[29]. For example, they can be used to transport information encoded on the phonon chirality or magnetic moment, similar to the role played by the electron spin/valley in spintronic/valleytronic devices. In experiments, chiral phonons have been detected indirectly through their interaction with optically bright excitons in two-dimensional semiconductors[30-35]. It has been pointed out that chiral phonons of electron crystals can couple efficiently to external fields as well as magnetic spin orders through the spin-orbit interaction, due to the small effective mass of the crystal sites[27]. This can facilitate the controlled generation, manipulation and direct observation of chiral phonons through their finite angular momenta and magnetizations, implying that electron crystals in bilayer TMDs can serve as an exciting new platform to study chiral phonons.

In this work, we theoretically analyze chiral phonons in the recently observed honeycomb-type bilayer Wigner crystals, taking into account the difference between the intralayer and interlayer Coulomb interactions due to the vertical layer separation and wide tunability of TMDs bilayers. The rest of the work is organized as follows: Section II gives the phonon dispersions of honeycomb-type bilayer Wigner crystals and discusses their dynamical stability; Section III analyzes the phonon chirality from the inversion and/or time-reversal symmetry breaking; Section IV is a brief summary.

## II. Phonon dispersions in honeycomb-type bilayer Wigner crystals

We consider bilayer Wigner crystals in TMDs/hBN/TMDs van der Waals structures with equal doping densities in the two TMDs layers, with hBN standing for hexagonal boron nitride layers. Such structures are usually encapsulated by thick hBN layers in experiments. In such a system the doped electrons can crystallize into a honeycomb-type Wigner crystal under a low temperature, with the **A** and **B** sublattice sites located in opposite TMDs layers (denoted as layer-**A** and layer-**B**, see Fig. 1(a)). We write the interlayer distance as $D$, and the Wigner crystal wavelength as $\lambda$. The in-plane equilibrium positions of electrons are denoted as $\boldsymbol{R}_{A,n} \equiv \boldsymbol{R}_n$ and $\boldsymbol{R}_{B,n} \equiv \boldsymbol{R}_n - \boldsymbol{R}_{AB}$ for **A** and **B** sites, respectively, with $n$ the index of the primitive cell. Two electrons in the same layer (opposite layers) with an in-plane displacement $\boldsymbol{R}$ interacts through the intralayer Coulomb potential $V_{\mathrm{intra}}(\boldsymbol{R})$ (interlayer Coulomb potential $V_{\mathrm{inter}}(\boldsymbol{R})$), which is affected by the 2D screening effect of the atomically thin layers. Here we treat the hBN spacer between the two TMDs layers and the encapsulating hBN as the dielectric environment with a dielectric constant $\varepsilon \approx 4.5$ (the average dielectric constant of bulk hBN). The dielectric screening of monolayer TMDs is characterized by its screening length $r_0 \approx 4.5/\varepsilon$ nm [36,37]. Following Ref. [38,39], we use the below forms for the Coulomb potentials

$$V_{\mathrm{intra}}(\boldsymbol{R}) = \frac{1}{\varepsilon} \int_0^\infty \frac{J_0(qR)}{1 + 2r_0 q}\,dq,$$
$$V_{\mathrm{inter}}(\boldsymbol{R}) = \frac{1}{\varepsilon} \int_0^\infty \frac{e^{-qD} J_0(qR)}{1 + 2r_0 q}\,dq. \quad (1)$$

Here $J_0$ is the Bessel function of the first kind. We note that more complicated Coulomb potential forms have been used in other literatures[40,41], which lead to quantitative different Coulomb potential strengths. Such changes, however, should not affect our qualitative conclusions about chiral phonons.

The vibration of electrons leads to a small deviation of the in-plane separation from the equilibrium value $\boldsymbol{R}_n$ or $\boldsymbol{R}_n + \boldsymbol{R}_{AB}$. The corresponding Coulomb interactions with $r \ll \lambda$ can be expanded as

$$\begin{aligned}
V_{\mathrm{intra}}(\boldsymbol{R}_n + \boldsymbol{r}) &\approx V_{\mathrm{intra}}(\boldsymbol{R}_n) \\
&+ \boldsymbol{r} \cdot \frac{\partial V_{\mathrm{intra}}(\boldsymbol{R}_n)}{\partial \boldsymbol{R}_n} + \frac{1}{2}\left(\boldsymbol{r} \cdot \frac{\partial}{\partial \boldsymbol{R}_n}\right)^2 V_{\mathrm{intra}}(\boldsymbol{R}_n), \\
V_{\mathrm{inter}}(\boldsymbol{R}_n + \boldsymbol{R}_{AB} + \boldsymbol{r}) &\approx V_{\mathrm{inter}}(\boldsymbol{R}_n + \boldsymbol{R}_{AB}) \\
&+ \boldsymbol{r} \cdot \frac{\partial V_{\mathrm{inter}}(\boldsymbol{R}_n + \boldsymbol{R}_{AB})}{\partial \boldsymbol{R}_n} + \frac{1}{2}\left(\boldsymbol{r} \cdot \frac{\partial}{\partial \boldsymbol{R}_n}\right)^2 V_{\mathrm{inter}}(\boldsymbol{R}_n + \boldsymbol{R}_{AB}).
\end{aligned} \quad (2)$$

We also consider a TMDs/TMDs/hBN/TMDs van der Waals structure, where the two TMDs layers in direct contact form a triangular-type moiré pattern. A moiré potential $U_{\mathrm{moiré}}$ can thus apply on electrons in layer-**A**, but not on those in layer-**B**, see Fig. 1(b). We assume that the moiré wavelength $\lambda$ is the same as that of the Wigner crystal. Near the equilibrium position $\boldsymbol{R}_n$, $U_{\text{moiré}}$ can be approximated by a harmonic trap

$$U_{\text{moiré}}(\boldsymbol{R}_n + \boldsymbol{r}) \approx \frac{\gamma}{2\lambda^2} r^2. \tag{3}$$

Here $\gamma$ is the moire confinement strength, which is expected to be in the order of ~ 0.1-1 eV[27].

Below we analyze the collective vibration or phonon modes of honeycomb-type bilayer Wigner crystals under various system parameters. We also consider the effect of an externally applied out-of-plane magnetic field $B$, as the small electron effective mass (comparable to the free electron mass $m_0$) results in the sensitive dependence of its motion on the magnetic field. Following the standard treatment of the lattice phonon[23-25,27], the frequency $\omega_{l,\boldsymbol{q}}$ of $l$-th phonon branch as a function of the wave vector $\boldsymbol{q}$ can be obtained from the below dynamical equation:

$$[\vec{\text{g}}(\omega_{l,\boldsymbol{q}} B) - \omega_{l,\boldsymbol{q}}^2] \vec{\psi}_{l,\boldsymbol{q}} = 0, \tag{4}$$

Here we use $l = 1$ to 4 to denote phonon branches from low to high frequencies. The dynamical matrix $\vec{\text{g}}(\omega_{l,\boldsymbol{q}} B)$ has the form

$$\vec{\text{g}}(\omega_{l,\boldsymbol{q}} B) = \begin{bmatrix} \frac{g + \frac{\gamma}{\lambda^2} - \omega_{l,\boldsymbol{q}} B}{m_A} & \frac{g_+}{m_A} & \frac{g^{AB}}{\sqrt{m_A m_B}} & \frac{g_+^{AB}}{\sqrt{m_A m_B}} \\ \frac{g_-}{m_A} & \frac{g + \frac{\gamma}{\lambda^2} + \omega_{l,\boldsymbol{q}} B}{m_A} & \frac{g_-^{AB}}{\sqrt{m_A m_B}} & \frac{g^{AB}}{\sqrt{m_A m_B}} \\ \frac{g^{BA}}{\sqrt{m_A m_B}} & \frac{g_+^{BA}}{\sqrt{m_A m_B}} & \frac{g - \omega_{l,\boldsymbol{q}} B}{m_B} & \frac{g_+}{m_B} \\ \frac{g_-^{BA}}{\sqrt{m_A m_B}} & \frac{g^{BA}}{\sqrt{m_A m_B}} & \frac{g_-}{m_B} & \frac{g + \omega_{l,\boldsymbol{q}} B}{m_B} \end{bmatrix}, \tag{5}$$

where $m_A$ ($m_B$) is the effective mass of electrons at **A** (**B**) sites, and

$$\begin{aligned} g &\equiv \frac{1}{2} \sum_{\boldsymbol{R}_n \neq 0} (1 - \cos(\boldsymbol{q} \cdot \boldsymbol{R}_n)) \nabla^2 V_{\text{intra}}(\boldsymbol{R}_n + \boldsymbol{r})|_{\boldsymbol{r}=0} + \frac{1}{2} \sum_{\boldsymbol{R}_n} \nabla^2 V_{\text{inter}}(\boldsymbol{R}_n + \boldsymbol{r})|_{\boldsymbol{r}=\boldsymbol{R}_{AB}}, \\ g_\pm &\equiv \frac{1}{2} \sum_{\boldsymbol{R}_n \neq 0} (1 - \cos(\boldsymbol{q} \cdot \boldsymbol{R}_n)) \partial_\pm^2 V_{\text{intra}}(\boldsymbol{R}_n + \boldsymbol{r})|_{\boldsymbol{r}=0} + \frac{1}{2} \sum_{\boldsymbol{R}_n} \partial_\pm^2 V_{\text{inter}}(\boldsymbol{R}_n + \boldsymbol{r})|_{\boldsymbol{r}=\boldsymbol{R}_{AB}}, \\ g^{AB} &\equiv \frac{1}{2} \sum_{\boldsymbol{R}_n} e^{i\boldsymbol{q} \cdot (\boldsymbol{R}_n + \boldsymbol{R}_{AB})} \nabla^2 V_{\text{inter}}(\boldsymbol{R}_n + \boldsymbol{r})|_{\boldsymbol{r}=\boldsymbol{R}_{AB}} = (g^{BA})^*, \\ g_\pm^{AB} &\equiv \frac{1}{2} \sum_{\boldsymbol{R}_n} e^{i\boldsymbol{q} \cdot (\boldsymbol{R}_n + \boldsymbol{R}_{AB})} \partial_\pm^2 V_{\text{inter}}(\boldsymbol{R}_n + \boldsymbol{r})|_{\boldsymbol{r}=\boldsymbol{R}_{AB}} = (g_\mp^{BA})^*, \end{aligned} \tag{6}$$

with $\nabla^2 \equiv \partial_x^2 + \partial_y^2$ and $\partial_\pm \equiv \partial_x \pm i\partial_y$. Here the summation involving $V_{\text{intra}}$ with a finite $r_0$ can be calculated using a modified Ewald method developed in Ref. [27].

In Eq. (4), the four-component vector $\vec{\psi}_{l,\boldsymbol{q}} \equiv [\sqrt{m_A} A_{l,\boldsymbol{q}}^+, \sqrt{m_A} A_{l,\boldsymbol{q}}^-, \sqrt{m_B} B_{l,\boldsymbol{q}}^+, \sqrt{m_B} B_{l,\boldsymbol{q}}^-]^T$ describes the circular polarization of the corresponding phonon mode, with $A_{l,q}^+$ and $A_{l,q}^-$ ($B_{l,q}^+$ and $B_{l,q}^-$) the left- and right-handed circular vibration amplitudes, respectively, for electrons at **A** (**B**) sites. The angular momentum of the phonon is thus proportional to $\left(m_A|A_{l,q}^+|^2 + m_B|B_{l,q}^+|^2\right)\omega_{l,q}^2 - \left(m_A|A_{l,q}^-|^2 + m_B|B_{l,q}^-|^2\right)\omega_{l,q}^2$. We can then define the phonon chirality as

$$P_{l,q} \equiv \frac{m_A\left(|A_{l,q}^+|^2 - |A_{l,q}^-|^2\right) + m_B\left(|B_{l,q}^+|^2 - |B_{l,q}^-|^2\right)}{m_A\left(|A_{l,q}^+|^2 + |A_{l,q}^-|^2\right) + m_B\left(|B_{l,q}^+|^2 + |B_{l,q}^-|^2\right)}. \tag{7}$$

$P_{l,q} \in [-1,1]$ can be viewed as the normalized angular momentum of the phonon. Since the crystal site carries a charge, its circular vibration can give rise to an experimentally observable magnetization proportional to $P_{l,q}$.

We first analyze the phonon dispersion $\omega_{l,q}$ and the dynamical stability of the bilayer Wigner crystal when the magnetic field is absent. It is known that in the monolayer limit ($D = 0$) without the moiré potential ($\gamma = 0$), the Wigner crystal can only be in the triangular lattice form and the honeycomb-type Wigner crystal is dynamical unstable[27]. With the increase of the interlayer distance $D$, $V_{\text{inter}}$ weakens whereas $V_{\text{intra}}$ is not affected. In the large $D$ limit, electrons in two layers will form two triangular lattices, which are interlocked by the residue interlayer Coulomb interaction to form a honeycomb-type bilayer Wigner crystal. We thus expect that the honeycomb-type Wigner crystal is stable only above a critical interlayer distance $D_c$. The dynamical stability of the honeycomb-type Wigner crystal can be analyzed from the calculated phonon dispersion, which requires all phonon frequencies to have real values (i.e., $\omega_{l,q}^2 > 0$). Fig. 1(c) and 1(d) shows our calculated phonon dispersions of honeycomb-type bilayer Wigner crystals with several different $D$ values, all under $m_A = m_B = 0.5 m_0$, $\lambda = 5$ nm and $\gamma = 0$. For $D = 1$ nm, the whole lowest-energy branch has imaginary $\omega_{l,q}$ values, indicating that the Wigner crystal is dynamically unstable against in-plane site fluctuations. When the interlayer distance increases, a crossover from imaginary to real for $\omega_{l,q}$ happens first near the high-symmetry point **K**, and then at other wave vectors, see the dispersion under $D = 1.2$ nm in Fig. 1(c) as an example. With the further increase of $D$, $\omega_{l,q}$ is imaginary only in a small region near **Γ** when $D$ reaches 2.1 nm, and becomes fully real under $D = 2.5$ nm (see Fig. 1(d)). This implies that above a critical value $D_c$, all phonon frequencies become real and the honeycomb-type bilayer Wigner crystal is dynamically stable. The crystal stability can be further enhanced by adding a moiré potential with $\gamma \neq 0$. Fig. 1(e) shows the phonon dispersions under $\lambda = 5$ nm, $D = 1$ nm, $\gamma = 0.15$ and 0.1 eV. $\omega_{l,q}$ is imaginary only in a small region near **M** under $\gamma = 0.1$ eV, but becomes fully real under $\gamma = 0.15$ eV. It should be emphasized that in this case the lowest-frequency of the phonon is located at **M**, which is different from Fig. 1(d) with $\gamma = 0$ where the lowest-frequency is at **Γ**. In Fig. 1(f) we summarize the calculated critical interlayer distance $D_c$ as a function of the 2D screening length $r_0$, at several different values of $\lambda$ and $\gamma$. Under a fixed $\lambda$, $D_c$ decreases with the increase of $r_0$ and $\gamma$.

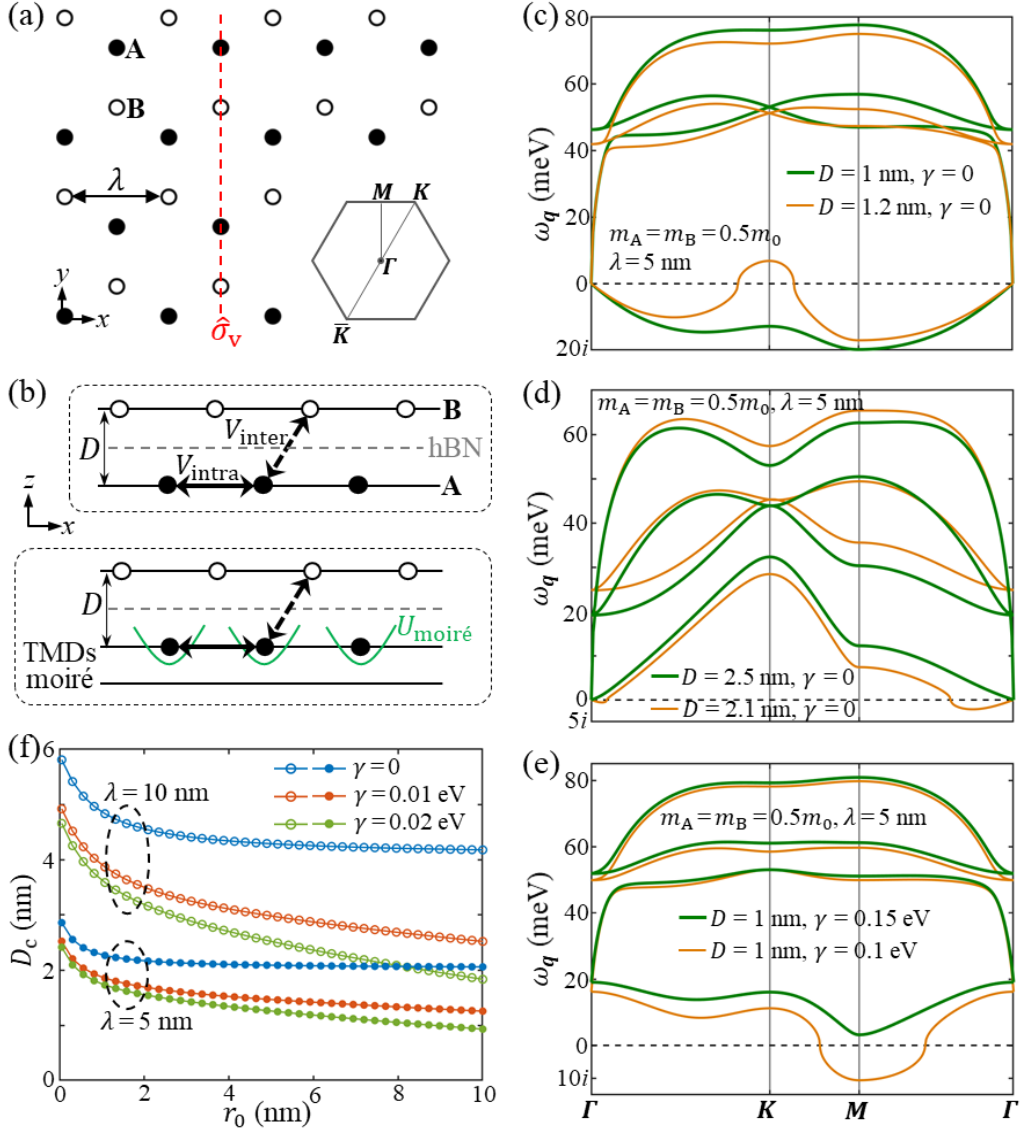

Fig. 1 (a) The top view of a honeycomb-type bilayer Wigner crystal with lattice constant $\lambda$. Electrons at **A** and **B** sublattice sites are denoted as solid and empty dots, respectively. The red dashed line denotes the vertical mirror plane $\hat{\sigma}_v$ of the crystal under the absence of a magnetic field. The inset shows the Brillouin zone with high-symmetry points $\Gamma$, $K$, $M$ and $\overline{K}$. (b) The top pattern shows a side view of the crystal in a TMDs/hBN/TMDs van der Waals structure, with A and B sites located in opposite layers. $D$ is the interlayer distance, $V_{intra}$ ($V_{inter}$) is the Coulomb interaction between electrons in the same layer (opposite layers). The bottom pattern is a side view of the crystal in a TMDs/TMDs/hBN/TMDs structure, with the two TMDs layers in direct contact forming a triangular-type moiré pattern. Electrons in the moiré pattern thus experience a moiré potential $U_{moiré}$ (green curves). (c) The calculated phonon dispersions under $\lambda = 5$ nm, $D = 1$ and 1.2 nm, with the absence of the moiré potential ($\gamma = 0$). (d) The phonon dispersions under $\lambda = 5$ nm, $D = 2.1$ and 2.5 nm and $\gamma = 0$. (e) The phonon dispersions under finite moiré potentials $\gamma = 0.1$ and 0.15 eV, when $\lambda = 5$ nm and $D = 1$ nm. (f) The critical interlayer distance $D_c$ as a function of the screening length $r_0$, under several values of wavelength $\lambda$ and moiré confinement strength $\gamma$. The effective masses in (c-f) are $m_A = m_B = 0.5m_0$.

### III. Chiral phonons under time-reversal or/and inversion symmetry breaking

Next we analyze the phonon chirality of the honeycomb-type bilayer Wigner crystal. In TMDs bilayers with $m_A = m_B$, $\gamma = 0$ and $B = 0$, the angular momentum or chirality vanishes for every non-degenerate phonon mode because of the parity-time symmetry. Setting $m_A \neq m_B$ and/or $\gamma \neq 0$ break the inversion symmetry, applying an out-of-plane magnetic field ($B \neq 0$) breaks the time-reversal symmetry, both can give rise to the emergence of chiral phonons. The $2\pi/3$-rotational ($\hat{C}_3$) symmetry of **K** and **Γ** wave vector positions implies that the corresponding vibrational patterns of **A** and **B** sites are both circularly polarized for non-degenerate phonons, where maximum chirality magnitudes ±1 can be reached. Thus, wave vectors near **K** and **Γ** are the focus of our analysis.

Fig. 2(a-d) show the phonon dispersion and chirality when the system is inversion asymmetric but time-reversal symmetric, where $\omega_{l,q} = \omega_{l,-q}$ and $P_{l,q} = -P_{l,-q}$. Note that the two intermediate phonon branches ($l = 2$ and 3) at **K** are doubly degenerate when the inversion symmetry is present, as illustrated in Fig. 1(c,d). Fig. 2(a) indicates that breaking the inversion symmetry by a small mass imbalance ($m_A = 0.5 m_0$ and $m_B = 0.6 m_0$) lifts this degeneracy slightly. The resultant splitting $\Delta\omega_{23,K}$ combined with the $\hat{C}_3$ symmetry at **K** leads to the fully chiral nature of the two intermediate branches, with $|P_{2,K}| = |P_{3,K}| = 1$ and $P_{2,K} = -P_{3,K}$. The values of $P_{l,q}$ along the path $\boldsymbol{\Gamma K M \Gamma \bar{K}}$ are shown as curves in Fig. 2(b), indicating that $|P_{l,q}|$ decays fast when moving away from **K**. Note that $P_{l,q} = 0$ for **q** along the **MΓ** line, which originates from the presence of $\hat{\sigma}_v$ symmetry in the Wigner crystal when $B = 0$ (see Fig. 1(a)). The lowest-energy ($l = 1$) and highest-energy ($l = 4$) branches at **K** are also chiral, but with small chirality values. Fig. 2(c) and 2(d) show the results under a large mass imbalance ($m_A = 0.5 m_0$ and $m_B = 2 m_0$), where the two intermediate branches have a large splitting at **K** with $\omega_{2,K}$ close to $\omega_{1,K}$ and $\omega_{3,K}$ close to $\omega_{4,K}$. In fact, the large mass imbalance leads to nearly decoupled vibrations of the two layers. We show the phonon dispersion under $V_{\text{inter}} = 0$ as dashed black curves in Fig. 2(c), which is simply the superposition of phonon branches from the two monolayer Wigner crystals. The calculated phonon dispersion under $V_{\text{inter}} \neq 0$ is close to the black curves, implying that the **A** and **B** sites in opposite layers are nearly decoupled. Note that the two phonon branches of each monolayer Wigner crystal are degenerate at **K**. When taking into account the residue effect of $V_{\text{inter}}$, the degeneracy is lifted and chiral phonons emerge at **K**. Now $P_{1,K} = -P_{4,K} \approx -1$ and $P_{2,K} = -P_{3,K} = +1$, see Fig. 2(d). Introducing a moiré potential with $\gamma \neq 0$ also breaks the inversion symmetry, which plays a similar role as the mass imbalance. When both the mass imbalance and moiré potential are present, their interplay can quantitatively change the phonon chirality values near **K**. Since the moiré potential is applied on **A** sites only, it always enhances the splitting $\Delta\omega_{23,K}$ between the two intermediate branches when $m_A < m_B$ (Fig. 2(e)). For the $m_A > m_B$ case, when increasing $\gamma$ from 0, $\Delta\omega_{23,K}$ decreases first (Fig. 2(f)) and becomes zero at a critical value $\gamma_c$. After reaching $\gamma_c$, $\Delta\omega_{23,K}$ again increases with $\gamma$. The signs of $P_{2,K}$ and $P_{3,K}$ are switched when changing $\gamma$ from $< \gamma_c$ to $> \gamma_c$.

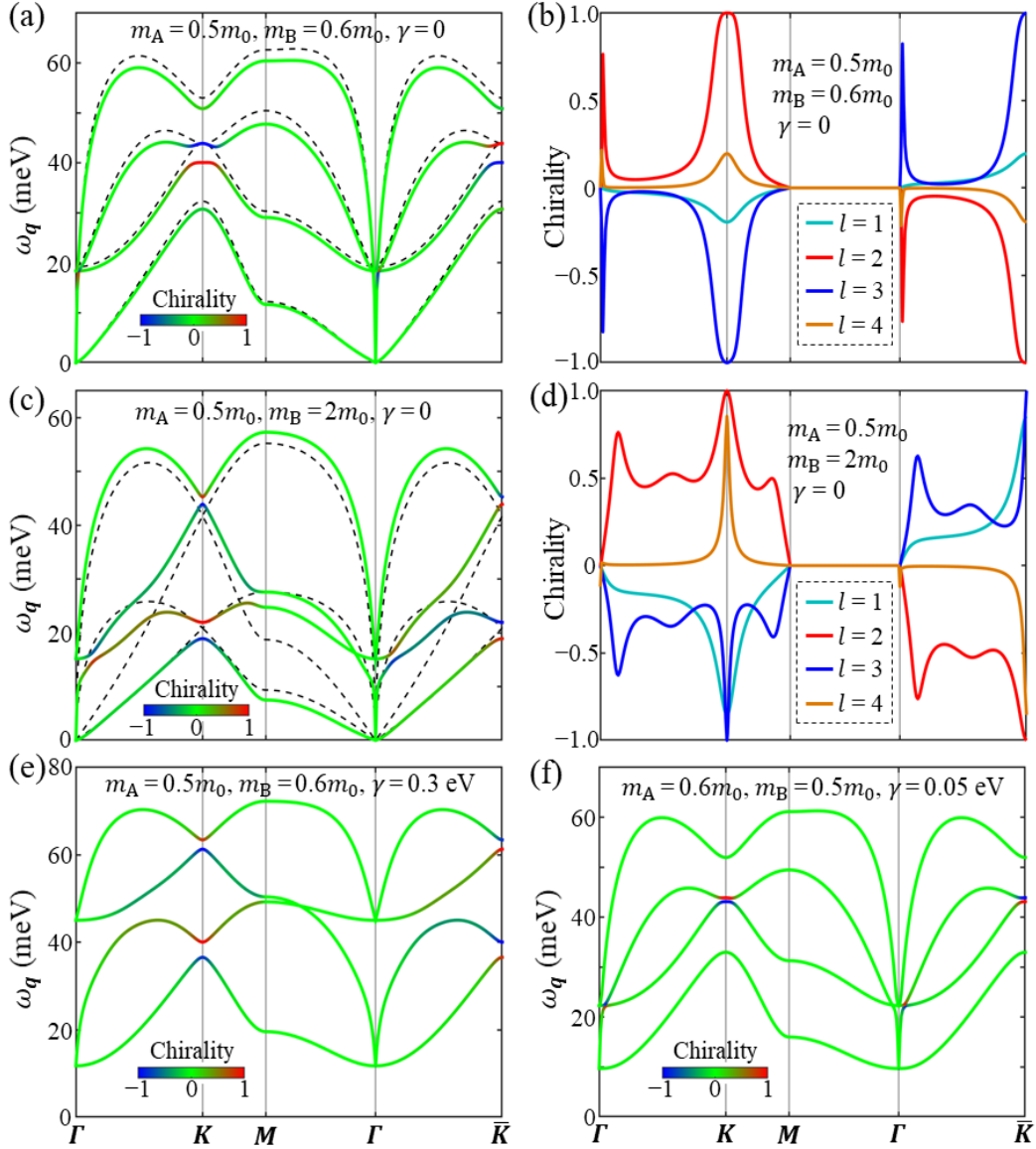

Fig. 2 (a) Black dashed lines are the phonon dispersion of an inversion symmetric bilayer Wigner crystal with $m_A = m_B = 0.5m_0$ and $\gamma = 0$. Colored lines show the phonon dispersion and chirality under a weak mass imbalance ($m_A = 0.5m_0$, $m_B = 0.6m_0$) and $\gamma = 0$, with the color denoting the chirality value. (b) The chirality values $P_{l,q}$ in (a) shown as curves. (c) Black dashed lines are the phonon dispersion of two decoupled monolayer Wigner crystals obtained from artificially setting $V_{inter} = 0$, under a large mass imbalance ($m_A = 0.5m_0$, $m_B = 2m_0$) and $\gamma = 0$. Colored lines show the corresponding phonon dispersion and chirality when the two layers are coupled by the finite $V_{inter}$. (d) The chirality values $P_{l,q}$ in (c) shown as curves. (e) The phonon dispersion and chirality under $m_A = 0.5m_0$, $m_B = 0.6m_0$ and a finite moiré potential $\gamma = 0.3$ eV. (f) The phonon dispersion and chirality under $m_A = 0.6m_0$, $m_B = 0.5m_0$ and a weak moiré potential $\gamma = 0.05$ eV. Other parameters in (a-f) are $\lambda = 5$ nm and $D = 2.5$ nm.

For long-wavelength phonons with $q$ close to $\Gamma$, they correspond to transverse acoustic (TA), longitudinal acoustic (LA), transverse optical (TO) and longitudinal optical (LO) modes from low to high frequencies. The acoustic and optical modes are doubly degenerate at $\Gamma$ due to the time reversal symmetry. Under $\gamma = 0$, the long-wavelength TA mode has a standard dispersion relation $\omega_{1,q} \propto q$, whereas the dispersion of the long-wavelength LA mode is $\omega_{2,q} \propto \sqrt{q}$. This can be understood from the following analysis. When $\gamma = 0$, the center-of-mass motion of the whole Wigner crystal is in a plane-wave form, resulting in a homogeneous charge distribution for the ground state without phonons. The vibrational pattern of the long-wavelength LA mode introduces a macroscopic charge modulation, which can be viewed as a plasmon oscillation in a homogeneous charge background. Thus, $\omega_{2,q}$ is simply the frequency of the 2D plasmon oscillation determined by Maxwell's equations. On the other hand, the long-wavelength TA mode doesn't introduce charge modulations, which thus has a standard dispersion. The above analysis doesn't apply to the $\gamma \neq 0$ case, as the Wigner crystal sites are now pinned to the moiré potential minima and the ground state charge distribution is no longer homogeneous. Although the two lowest-frequency modes have $\omega_{1,\Gamma} = \omega_{2,\Gamma} > 0$ under $\gamma \neq 0$, but we still denote them as TA and LA since they correspond to all sites oscillating in phase. The dispersion relations now become $\omega_{1,q} - \omega_{1,\Gamma} \propto q^2$ and $\omega_{2,q} - \omega_{2,\Gamma} \propto q$. Due to the fast increase of $\omega_{2,q}$ with $q$, avoided crossings between LA and optical modes can occur near $\Gamma$, resulting in significant $P_{l,q}$ values for the corresponding phonon modes (see Fig. 2(b,d)). Such chiral phonons, however, are accidental and not protected by symmetries. The corresponding $q$ positions and chirality values depend sensitively on system parameters, thus are not the focus of this work.

Below we analyze the effect of applying an out-of-plane magnetic field $B$ on an inversion symmetric bilayer Wigner crystal with $m_A = m_B = 0.5 m_0$ and $\gamma = 0$. Fig. 3(a) shows the result under a relatively weak magnetic field $B = 10$ T. Compared to the inversion asymmetric but time-reversal symmetric case in Fig. 2, Fig. 3(a) shows similar dispersion and chirality patterns near $K$ with $|P_{2,K}| = |P_{3,K}| = 1$. The most significant differences are: (1) the emergence of finite $P_{l,q}$ values for $q$ along the $M\Gamma$ line, which is a result from the breaking of $\hat{\sigma}_v$ symmetry by the magnetic field; (2) the degeneracy lifting and emergence of chiral phonons at $\Gamma$, where the two acoustic/optical modes exhibit a splitting $B/m_A$ ($\approx 2$ meV under $B = 10$ T and $m_A = 0.5 m_0$). Fig. 3(b) is an enlarged view of the phonon dispersion and chirality near $\Gamma$, which indicates that the three branches with $\omega_{l,\Gamma} > 0$ now become chiral. The absolute values of the chirality are $|P_{2,\Gamma}| = |P_{3,\Gamma}| = |P_{4,\Gamma}| = 1$, which decay to $\approx 0$ when slightly away from $\Gamma$ (Fig. 3(c)). $P_{1,q}$ of the lowest-frequency branch, however, is close to zero in the entire Brillouin zone. The result under a strong magnetic field $B = 70$ T is presented in Fig. 3(d), and Fig. 3(e) corresponds to its enlarged view near $\Gamma$. Fig. 3(f) shows the $P_{l,q}$ values in Fig. 3(d) as curves. Compared to Fig. 3(c) under $B = 10$ T, we can see that the strong field value substantially changes the chirality. In Fig. 3(f), $|P_{l,q}|$ values for $l = 2, 3$ and $4$ are large in the entire Brillouin zone. For the lowest-frequency branch, $P_{1,q}$ also becomes rather significant near $K$ and $\bar{K}$, but $\omega_{1,q} \approx 0$ and $P_{1,q} \approx 0$ for $q$ near $\Gamma$ are not affected even when the magnetic field is very strong.

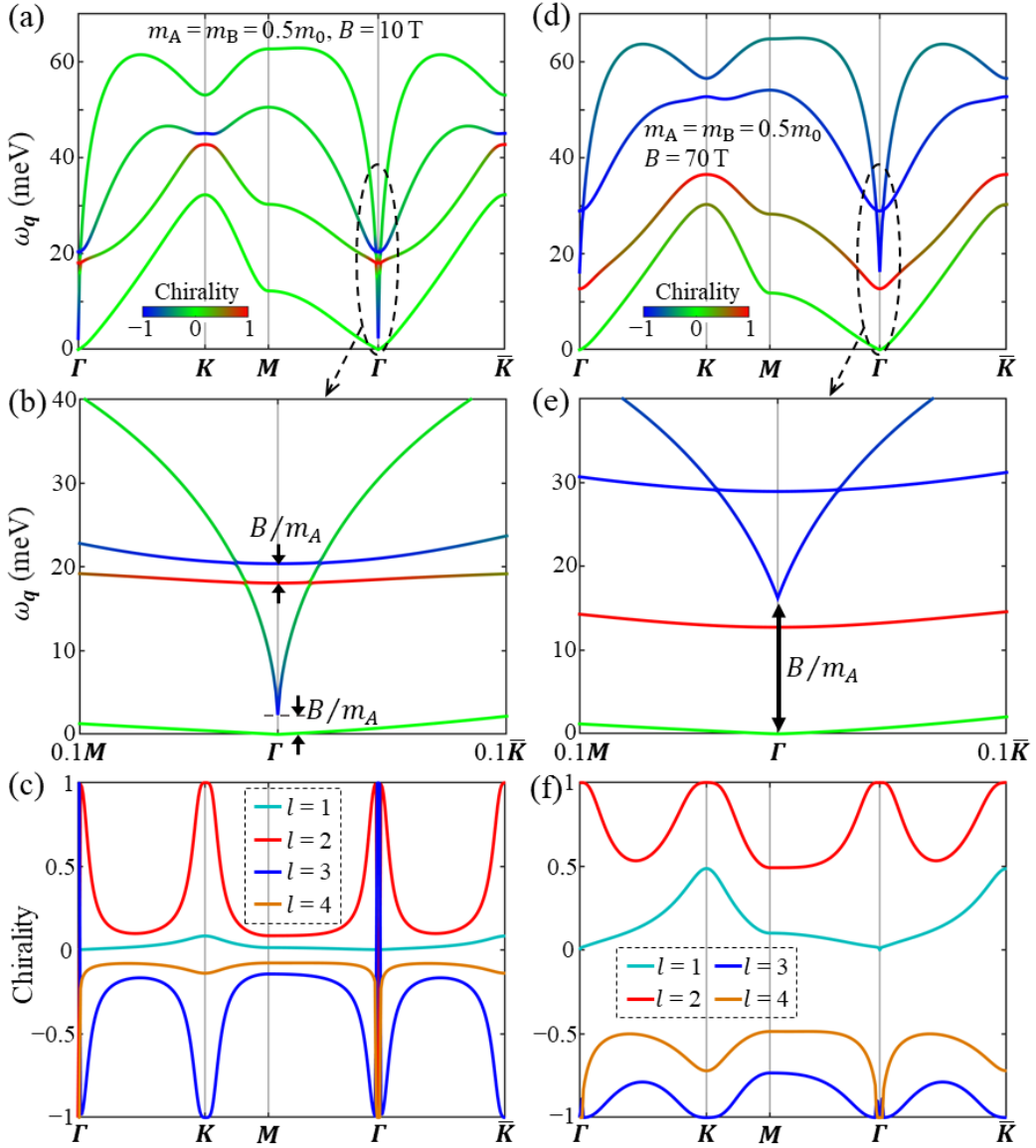

Fig. 3 (a) The phonon dispersion and chirality of an inversion symmetric bilayer Wigner crystal under a relatively weak magnetic field $B = 10$ T. (b) An enlarged view of (a) near $\Gamma$. The magnetic field introduces a splitting $B/m_A$ to the phonon frequencies at $\Gamma$. (c) The chirality values in (a) for $q$ along the path $\Gamma K M \Gamma \bar{K}$. (d) The phonon dispersion and chirality under a strong magnetic field $B = 70$ T. (e) An enlarged view of (d) near $\Gamma$. (f) The chirality values $P_{l,q}$ in (d). Other parameters in (a-f) are $\lambda = 5$ nm, $D = 2.5$ nm, $m_A = m_B = 0.5 m_0$ and $\gamma = 0$.

When the system lacks both the inversion and time-reversal symmetry, phonons with wave vectors $q$ and $-q$ are no longer related by $\omega_{l,q} = \omega_{l,-q}$ and $|P_{l,q}| = |P_{l,-q}|$. Below we focus on the case that the inversion symmetry breaking is induced by $\gamma \neq 0$, as when $\gamma = 0$ and $B \neq 0$ the phonon chirality under $m_A \neq m_B$ is qualitatively the same as those with $m_A = m_B$. Fig. 4(a) shows the phonon dispersion and chirality under $m_A = m_B = 0.5 m_0$, $\gamma = 0.1$ eV and $B = 10$ T, and an enlarged view near $\Gamma$ is presented in Fig. 4(b). Compared to Fig. 3(a,b) with $\gamma = 0$, the most significant change in Fig. 4(a,b) is the emergence of the chirality $P_{1,\Gamma} = +1$ for the TA mode. Note that all long-wavelength LA, TO and LO modes become chiral when $B \neq 0$, which can be understood as a result from the strong coupling between the introduced spatial charge modulations and the magnetic field. On the contrast, long-wavelength TA modes under $\gamma = 0$ are always non-chiral, as they don't introduce charge modulations. However, the ground state charge distribution is not homogeneous when $\gamma \neq 0$, and vibrational patterns of TA modes can also introduce spatial charge modulations. As a result, in Fig. 4(a,b) all four branches at $\Gamma$ are chiral with $P_{l,\Gamma}$ being either +1 or −1.

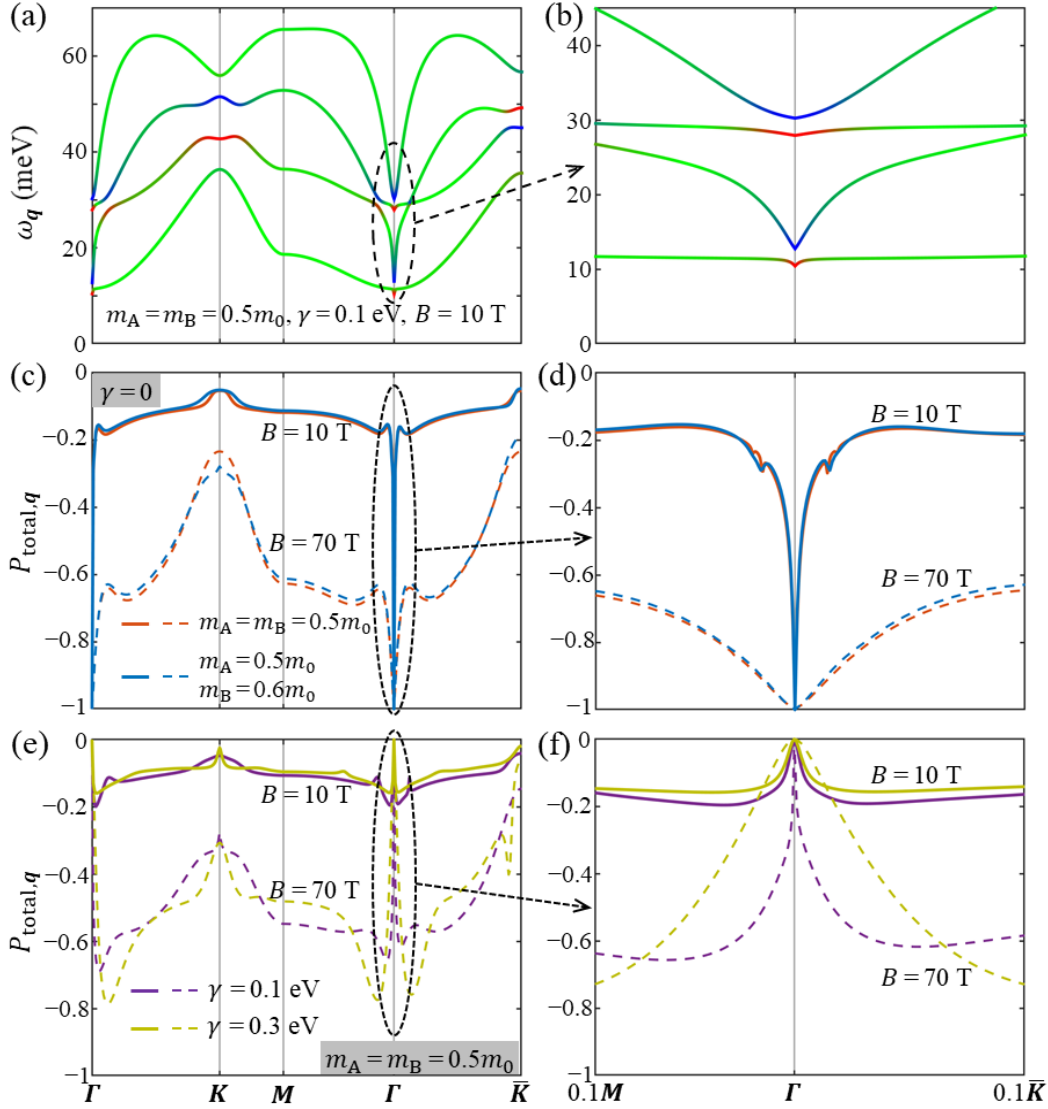

Fig. 4 (a) The phonon dispersion and chirality of a bilayer Wigner crystal without the inversion and time reversal symmetries, under $m_A = m_B = 0.5 m_0$, $\gamma = 0.1$ eV and $B = 10$ T. (b) An enlarged view of (a) near $\Gamma$. (c) The total chirality $P_{\text{total},q} \equiv \sum_{l=1}^{4} P_{l,q}$ of the four branches as a function of $q$ under $\gamma = 0$. (d) An enlarged view of (c) near $\Gamma$. (e) $P_{\text{total},q}$ under $\gamma \neq 0$. (f) An enlarged view of (e) near $\Gamma$.

Lastly, we consider the chirality sum $P_{\text{total},q} \equiv \sum_{l=1}^{4} P_{l,q}$ as a function of the wave vector $q$. Under $B = 0$, $P_{\text{total},q}$ vanishes for any $q$, because $\vec{\psi}_{l,q}$ and $\vec{\psi}_{l',q}$ with $l \neq l'$ correspond to different eigen-vectors of the same dynamical matrix $\overset{\leftrightarrow}{g}(\omega_{l,q} B = 0)$ thus are orthogonal to each other. However, $\vec{\psi}_{l,q}$ and $\vec{\psi}_{l',q}$ under $B \neq 0$ become generally non-orthogonal, since they now correspond to eigen-vectors of different dynamical matrices ($\vec{g}(\omega_{l,q}B)$ and $\vec{g}(\omega_{l',q}B)$). As a result, $P_{\text{total}}(q)$ is generally nonzero. Fig. 4(c) and 4(e) show our calculated $P_{\text{total}}(q)$ under $\gamma = 0$ and $\gamma \neq 0$, respectively. Away from $\Gamma$, $P_{\text{total},q}$ varies smoothly with $q$, whose absolute value increases with $B$. For the wave vector at $\Gamma$, $P_{\text{total},\Gamma} = -1$ for $\gamma = 0$ but $P_{\text{total},\Gamma} = 0$ for $\gamma \neq 0$, which originates from the distinct responses of the corresponding TA modes to the magnetic field. In the vicinity of $\Gamma$, $P_{\text{total},q}$ appears as a sharp dip for $\gamma = 0$ but a sharp peak for $\gamma \neq 0$, as can be seen from the enlarged views in Fig. 4(d) and 4(f), respectively. A stronger moiré confinement strength $\gamma$ leads to a larger width of the dip/peak. It should be noted that the above conclusion $P_{1,\Gamma} = 0$ under $\gamma = 0$ is based on the assumption that the center-of-mass motion of the whole Wigner crystal has the continuous translational invariance, which only applies to ideal cases without disorder potentials. In realistic experiments, Wigner crystal sites are always pinned due to the inevitable disorders even when the moiré potential is absent. In this case, TA modes near $\Gamma$ are also chiral under $B \neq 0$ and $\gamma = 0$.

## IV. Conclusion

In this work, we have analyzed the chiral phonons in honeycomb-type bilayer Wigner crystals, which can connect the two active research fields of strongly correlated physics and chiral phonons. The inversion symmetry of the system can be broken by introducing an effective mass imbalance to the two layers or a moiré potential to one layer, which can give rise to chiral phonons near $K$ and $\bar{K}$. Meanwhile, applying an out-of-plane magnetic field can break the time-reversal symmetry and result in the emergence of chiral phonons near $K$, $\bar{K}$ and $\Gamma$. Considering the wide tunability of the layered TMDs system, the dynamical stability of the bilayer Wigner crystal, the frequency and chirality of the chiral phonon can all be tuned by varying system parameters like the interlayer separation, mass imbalance, moiré confinement strength and magnetic field strength. These discoveries imply that honeycomb-type bilayer Wigner crystals can act as an exciting new platform with unprecedented opportunities to explore chiral phonons and the related phononic devices. For example, a topological transition can happen when the phonon bandgap at $K$ experiences a close-open-close process by tuning the external magnetic field strength. As a result, the phonon bands can become topologically nontrivial with nonzero Chern numbers[27], and topologically protected edge phonon modes can emerge. This can have potential applications in novel topological phononic devices[42].

**Acknowledgments:** This work was supported by Tencent's program of Aspiring Explorers in Science. H.Y. acknowledges support by NSFC under Grant No. 12274477 and the Department of Science and Technology of Guangdong Province in China (No. 2019QN01X061).